\documentclass[12pt]{article}
\usepackage{amsfonts}
\usepackage{amsmath}

\newcommand{\g}{{\cal G}}
\newcommand{\be}{\begin{equation}}
\newcommand{\ee}{\end{equation}}
\newcommand{\bea}{\begin{eqnarray}}
\newcommand{\eea}{\end{eqnarray}}

\newcommand{\ti}{{\mbox{i}}}

\newcommand{\Lg}{{\bf g}}

\newcommand{\del}{{\partial}}

\makeatletter

\@addtoreset{equation}{section}
\def\section{\@startsection {section}{1}{\z@}{-3.5ex plus -1ex minus
 -.2ex}{2.3ex plus .2ex}{\large\bf\centering}}
\def\subsection{\@startsection{subsection}{2}{\z@}{-3.25ex plus -1ex minus -.2ex}{1.5ex plus .2ex}{\bf}}
\def\subsubsection{\@startsection{subsubsection}{3}{\z@}{-3.25ex plus -1ex minus -.2ex}{1.5ex plus .2ex}{\sl}}
\makeatother
\begin{document}

\baselineskip 18pt \parindent 12pt \parskip 12pt

\begin{titlepage}

\begin{center}
{\Large {\bf Darboux transformation of the generalized coupled dispersionless integrable system }}\\\vspace{1in} {\large M. Hassan   \footnote{
mhassan@physics.pu.edu.pk, mhassan@maths.gla.ac.uk }
}\vspace{0.15in}

{{\it Department of Mathematics, University of Glasgow,\\
Glasgow G12 8QW, Scotland}}\\
{{\it and}}\\
{{\it Department of Physics, University of the Punjab,\\
Quaid-e-Azam Campus, Lahore-54590, Pakistan.}}\\

\end{center}

\vspace{1cm}
\begin{abstract}
The Darboux transformation on matrix solutions to the generalized coupled dispersionless integrable system based on some non-abelian Lie group, is studied and the solutions are shown to be expressed in terms of quasideterminants. As an explicit example, the Darboux transformation on scalar solutions to the system based on the Lie group $SU(2)$ is discussed in detail and the solutions are shown to be expressed as ratios of determinants.
\end{abstract}
\vspace{1cm} PACS: 02.30Zz, 02.30.Ik
\end{titlepage} 

\section{Introduction}

There has been a considerable interest in dispersionless integrable hierarchies for the last couple of decades, partly because of their emergence in diverse areas of mathematical and theoretical physics such as quantum field theory, conformal field theory, string theory, soliton theory etc. (see e.g. \cite{Kri}-\cite{dun}). These integrable systems arise as semi-classical limit of the ordinary integrable systems. In other words the quasi-classical limit corresponds to the solutions which slowly depend on the independent variables resulting in the elimination of dispersion term of the original integrable equation. In recent past, the coupled dispersionless integrable system and its generalization, has also attracted a great deal of interest because of its nice integrability structure and its soliton solutions \cite{konno}-\cite{ala1}. However, it should be mentioned here that the coupled dispersionless integrable system introduced in Refs. \cite{konno}-\cite{kaku}, is referred to as dispersionless in the sense that it does not contain dispersion term and not in the sense of quasi-classical limit of some ordinary integrable system where the dispersion term is eliminated by taking an appropriate limit. The coupled dispersionless integrable system and its generalization is solvable by the inverse scattering method, possesses infinite number of conservation laws and has Painleve property \cite{konno}-\cite{ala1}. For the case of Lie group $SU(2) $, the multi-soliton solutions have been investigated using Backlund and Darboux transformations \cite{ala2}-\cite{chen}.

In this paper, we study the Darboux transformation of the generalized coupled dispersionless integrable system based on some Lie group $\g $ and obtain its solutions expressed in terms of quasideterminants of Gelfand and Retakh \cite{gr}-\cite{gr3}. For the special case of $\g =SU(2) $, we show that the scalar solutions to the system are expressed as ratios of determinants showing the transition from the matrix case to the scalar case. We compare our results with those of \cite{chen} and also with the previously known results of the sine-Gordon system. For the construction of multi-solitons, we use the method of Darboux matrix and by iteration of Darboux transformation, we obtain the quasideterminant solutions to the generalized coupled dispersionless integrable system.

The layout of the paper is as follows. In section 2, we give a review of the generalized coupled dispersionless integrable system by writing its lagrangian, lagrange equations of motion and the Lax pair. In section 2, we also introduce the notion of quasideterminants which shall be used in section 3 to express the solutions. In section 3, we define a Darboux transformation via a Darboux matrix on matrix solutions to the generalized coupled dispersionless integrable system and express the solutions to the system in a closed form as quasideterminants. In section 4, we take an explicit example of the system when the underlying Lie group is $SU(2)$ and write its multisoliton solutions as ratios of determinants by using the Darboux transformation already discussed in section 3. At the end we compare our results with the already known results and make concluding remarks in the last section.

\section{The generalized coupled dispersionless integrable system}
The generalized coupled dispersionless integrable system based on some non-abelian Lie group $\g $ is a classical lagrangian field theory defined by the following action functional \cite{kaku1}
\begin{equation}
I=\int dt dx {\cal L}\left(S, \del_x S, \del_t S \right),
\end{equation}
with the lagrangian density defined by
\begin{equation}
{\cal L}={\rm Tr}\left(\frac{1}{2} \del_x S \del_t S - \frac{1}{3} G \left[S,\left[\del_x S, S\right]\right]\right), \label{cid}
\end{equation}
where $S$ is some matrix field and $G$ is a constant matrix; taking values in non-abelian Lie algebra $\Lg $ of the Lie group $\g $. Let us introduce a basis of anti-hermitian generators $T^a$ for
$\Lg$ with real, totally antisymmetric structure constants $f^{abc}$ and
normalization given by
\be
[ T^a , T^b ] = f^{abc} T^c \; , \qquad {\rm Tr} (T^a T^b) = -
\delta^{ab} \; .
\ee
For any $X \in \Lg$ we write
\be\label{alcpts}
X = T^{a} X^a \;  \qquad X^a = - {\rm Tr}(T^a X) \; .
\ee
The matrix fields $S$ and $G$ of the generalized coupled dispersionless system (\ref{cid} ) take values in the Lie algebra $\Lg $, therefore, we write
\begin{eqnarray}
S&=& \phi^a T^a ,\nonumber \\
G&=& \kappa^a T^a ,\nonumber
\end{eqnarray}
where $\phi=\phi(x,t)$ is a vector field with components $\{\phi^a , a=1,2,\cdots , {\rm dim} \,\Lg \}$ and $\kappa $ is a constant vector with components $\{\kappa^a , a=1,2,\cdots , {\rm dim} \, \Lg \}$.

The lagrange equation of motion resulting from the action is given by
\begin{equation}
\del_t \del_x S - [[S,G], \del_x S]=0. \label{eom}
\end{equation}
The field equation (\ref{eom}) is known as the generalized coupled dispersionless integrable system based on a Lie group $\g $. The equation (\ref{eom}) has an $n \times n $ AKNS (Ablowitz-Kaup-Newell-Segur) representation and the nonlinearity of the equation is due to the non-abelian character of the underlying Lie group. The equation (\ref{eom}) is a generalization of the coupled dispersionless equation originally introduced in \cite{konno} for $\g=SL(2, \mathbb{R})$ and $SU(2)$. For $\g=SL(2, \mathbb{R})$, the equation (\ref{eom}) reduces to
\begin{eqnarray}
\del_t \del_x q + \del_x (rs) &=&0, \nonumber \\
\del_t \del_x r-2 \del_x q r &=&0, \nonumber \\
\del_t \del_x s-2 \del_x q s &=&0, \nonumber
\end{eqnarray}
where $q, r $ and $s$ are real valued functions of $x$ and $t$. For $\g=SU(2)$, the system (\ref{eom}) is equivalent to
\begin{eqnarray}
\del_t \del_x q + r\del_x \bar{r} &=&0, \nonumber \\
\del_t \del_x r-2 \del_x q r &=&0, \nonumber \\
\del_t \del_x \bar{r}-2 \del_x q \bar{r} &=&0, \label{a}
\end{eqnarray}
where $r$ is a complex valued function and $\bar{r}$ denotes complex conjugation of $r$. The system (\ref{a}) is in fact equivalent to Pohlmeyer-Lund-Regge system \cite{kot}. If $r$ is real then the system (\ref{a}) is the one originally proposed in \cite{konno} and is equivalent to the sine-Gordon equation \cite{hirota}. With reference to Darboux transformation, we shall discuss it in detail in section 4.

The generalized coupled dispersionless integrable system (\ref{eom}) can be expressed as the compatibility condition of the following Lax pair \cite{kaku1}
\begin{eqnarray}
\del_x \psi&=&U(x,t,\lambda)\psi , \label{l1} \\
\del_t \psi&=&V(x,t,\lambda)\psi ,\label{l2}
\end{eqnarray}
where $\psi \in \g $, $\lambda$ is the spectral parameter and the matrix fields $U(x,t,\lambda)$ and $V(x,t,\lambda)$ are given by
\bea
 U(x,t,\lambda)&=&\lambda \del_x S , \notag \\
 V(x,t,\lambda)&=&[S,G]+\lambda^{-1} G . \notag
\eea
The compatibility condition of (\ref{l1})-(\ref{l2}) is the zero-curvature condition for the matrices $U(x,t,\lambda)$ and $V(x,t,\lambda)$
\be
\del_t U(x,t,\lambda)-\del_x V(x,t,\lambda)+[U(x,t,\lambda),V(x,t,\lambda)]=0 \label{zc},
\ee
which is equivalent to the equation of motion (\ref{eom}). In the next section, we define the Darboux transformation via a Darboux matrix on matrix solutions $\psi $ of the Lax pair (\ref{l1})-(\ref{l2}). To write down the explicit expressions for matrix solutions to the generalized coupled dispersionless integrable system, we will use the notion of quasideterminant introduced by Gelfand and Retakh \cite{gr}-\cite{gr3}.

Let $X$ be an $n \times n $ matrix over a ring $R$ (noncommutative, in general). For any $1\leq i $, $j\leq n $, let $r_i$ be the $i$th row and $c_j $ be the $j$th column of $X$. The matrix $X$ has $n^2 $ quasideterminants denoted by $|X|_{ij} $ for $i, j=1,\ldots , n $ and are defined by \be
|X|_{ij} =x_{ij} - r_i^{\,\, j} \left(X^{ij} \right)^{-1} c_j^{\,\, i},
\ee
where $x_{ij}$ is the $ij$th entry of $X$, $r_i^{\,\, j}$ represents the $i$th row of $X$ without the $j$th entry, $c_j^{\,\, i} $ represents the $j$th column of $X$ without the $i$th entry and $X^{ij}$ is the submatrix of $X$ obtained by removing from $X$ the $i$th row and the $j$th column. The quasideterminats are also denoted by the following notation
\be
|X|_{ij}= \left \vert \begin{array}{cc}
                        X^{ij} & c_j^{\,\, i} \\
                        r_i^{\,\, j} & \frame{\fbox{$ x_{ij} $}}
                      \end{array} \right \vert .
\ee
If the entries of the matrix $X$ all commute, then
\be
|X|_{ij}=(-1)^{i+j} \frac{{\rm det} X}{{\rm det} X^{ij}}.
\ee
For a detailed account of quasideterminants and their properties see e.g. \cite{gr}-\cite{gr3}. In this paper, we will consider only quasideterminants that are expanded about an $n \times n$ matrix over a commutative ring. In our case the ring $R$ is the (noncommutative) ring of $n \times n $ matrices over another commutative ring. Let
\be\left(
     \begin{array}{cc}
       A & B \\
       C & D \\
     \end{array}
   \right),\nonumber
\ee be a block decomposition of any $N \times N $ matrix where the
matrix $D$ is $n \times n$ and $A$ is invertible. The
quasideterminant expanded about the matrix $D$ is defined by \be
\left \vert \begin{array}{cc}
                        A & B \\
                        C & \frame{\fbox{$ D $}}
                      \end{array} \right \vert =D-CA^{-1} B .
\ee
\section{Darboux transformation}
The Darboux transformation on the matrix solutions of the Lax pair (\ref{l1})-(\ref{l2})
is defined in terms of an $n \times n$ matrix $D(x,t,\lambda)$, called the Darboux matrix.
The Darboux matrix connects the two matrix solutions of the Lax pair (\ref{l1})-(\ref{l2}),
such that the Lax pair is covariant under the Darboux transformation (see e.g. \cite{levi}-\cite{guhu}).
Let us denote the new matrix solution to the Lax pair (\ref{l1})-(\ref{l2}) by $\psi [1]$,
so that the Darboux transformation is defined by
\be
\psi[1]= D (x,t,\lambda ) \psi \label{dt-1}.
\ee
For the present case, we make the following ansatz for the Darboux matrix
\be
D(x,t,\lambda)= \lambda^{-1}I-  M (x, t),
\ee
where $M(x, t)$ is some $n \times n $ matrix field to be determined and $I$ is an $n \times n $
identity matrix. For the covariance of the Lax pair, we require that
\bea
\del_x  \psi[1] &=& \lambda \del_x S[1] \psi [1], \\
\del_t \psi[1] &=& [S[1], G[1]] \psi[1]+ \lambda^{-1} G[1] \psi[1],
\eea
where $S[1]$ and $G[1]$ are the Darboux transformed matrix fields defined by
\bea
S[1]&=& S- M, \nonumber \\
G[1]&=& G \quad ({\rm constant\,\, matrix}), \label{dt-2}
\eea
such that the matrix $M$ is required to satisfy the following conditions
\bea
\del_x M M&=& [\del_x S , M ] \label{m1}, \\
\del_t M &=& [[S, G], M]+[G, M]M .\label{m2}
\eea
The next step is to determine the matrix $M$ in terms of a particular solution of the Lax pair, such the $M $ satisfies the conditions (\ref{m1})-(\ref{m2}). For this pupose, we proceed as follows.

Let $\lambda_1 , \cdots , \lambda_n $ be non-zero, distinct real or complex constant parameters and $e_1 , \cdots ,e_n $ be constant column vectors such that
\be
\Theta = (\psi(\lambda_1 )e_1, \cdots , \psi(\lambda_n )e_n )=(\theta_1 , \cdots , \theta_n ),
\ee
be an invertible $n \times n$ matrix. Each column $\theta_i = \psi (\lambda_i )e_i $ of the matrix $\Theta $ is a column solution of the Lax pair (\ref{l1})-(\ref{l2}) at $\lambda = \lambda_i $, {\rm i.e. }the columns $\theta_i , \, i=1,\cdots , n $ satisfy
\bea
\del_x  \theta_i &=& \lambda_i \del_x S \theta_i , \\
\del_t \theta_i &=& [S, G] \theta_i+ \lambda_i^{-1} G \theta_i .
\eea
Let us define an invertible constant diagonal matrix with entries being the eigenvalues $\lambda_i $ corresponding to the eigenvectors $\theta_i $
\be
\Lambda = {\rm diag }(\lambda_1 , \cdots , \lambda_n ).
\ee
The Lax pair (\ref{l1})-(\ref{l2}) can now be written in the matrix form as
\bea
\del_x  \Theta &=& \del_x S  \Theta \Lambda ,\\
\del_t \Theta&=& [S, G] \Theta+  G \Theta \Lambda^{-1} ,
\eea
where $\Theta $ is a particular matrix solution of the Lax pair with a matrix $\Lambda $ of particular eigenvalues.

Now we show that the matrix $M= \Theta\Lambda^{-1} \Theta^{-1} $ expressed in terms of particular matrix solution $\Theta $ of the Lax pair (\ref{l1})-(\ref{l2}), satisfies the conditions (\ref{m1})-(\ref{m2}) on the matrix $M$ imposed by the covariance of the Lax pair under the Darboux transformation (\ref{dt-1}) and (\ref{dt-2}). For this purpose, let us take the $x$ derivative of the matrix $M= \Theta\Lambda^{-1} \Theta^{-1} $
\bea
\del_x M &=& \del_x \Theta \Lambda^{-1} \Theta^{-1} + \Theta \Lambda^{-1} \del_x \Theta^{-1} ,\nonumber \\
&=& \del_x S - \Theta \Lambda^{-1} \Theta^{-1} \del_x S \Theta \Lambda \Theta^{-1}, \nonumber \\
&=& \del_x S - M \del_x S M^{-1},
\eea
which is equation (\ref{m1}). Now take the $t$ derivative of $M= \Theta\Lambda^{-1} \Theta^{-1} $ to get
\bea
\del_t M &=& \del_t \Theta \Lambda^{-1} \Theta^{-1} + \Theta \Lambda^{-1} \del_t \Theta^{-1}, \nonumber \\
&=& [S,G]\Theta\Lambda^{-1} \Theta^{-1}+ G \Theta\Lambda^{-2} \Theta^{-1}- \Theta \Lambda^{-1} \Theta^{-1} [S, G]- \Theta \Lambda^{-1} \Theta^{-1} G \Theta \Lambda^{-1} \Theta^{-1}, \nonumber \\
&=&[S,G]M+GM^2 - M[S,G]-M G M ,\nonumber \\
&=&[[S,G],M ] +[G, M] M ,
\eea
which is equation (\ref{m2}). This shows that the matrix $M= \Theta\Lambda^{-1} \Theta^{-1} $ satisfies the conditions (\ref{m1})-(\ref{m2}). So we have established that the transformation
\be
\psi[1] = D(x,t, \lambda ) \psi = (\lambda^{-1} I - \Theta\Lambda^{-1} \Theta^{-1} ) \psi , \label{dt0}
\ee
constitutes the Darboux transformation on the matrix solution $\psi $ of the Lax pair (\ref{l1})-(\ref{l2}) of the generalized coupled dispersionless integrable system (\ref{eom}). The corresponding Darboux transformations on the matrix fields $S$ and $G$ are
\bea
S[1]&=& S- \Theta\Lambda^{-1} \Theta^{-1} ,\nonumber \\
G[1] &=& G . \label{dt1}
\eea
By introducing the notation $\psi^{(1)}=\lambda^{-1}\psi $ and $\Theta^{(1)}=\Theta\Lambda^{-1} $, the one-fold Darboux transformations (\ref{dt0})-(\ref{dt1}) can also be expressed in terms of quasideterminants as
\bea
\psi[1]&=&\left\vert
\begin{array}{cc}
\Theta  & \ \psi \\
\Theta^{(1)} & \frame{\fbox{$ \psi^{(1)} $}}%
\end{array}%
\right\vert ,\label{quasi1} \\
S[1]&=& S + \left\vert
\begin{array}{cc}
\Theta  & I \\
\Theta^{(1)}& \frame{\fbox{$ O $}}%
\end{array}%
\right\vert ,
\eea
where $O$ is an $n \times n$ null matrix.
The iteration of Darboux transformation $N$ times gives the quasideterminant matrix solution to the generalized coupled dispersionless integrable system (\ref{eom}). For each $k=1,2,\cdots , N $, let $\Theta_k $ be an invertible matrix solution of the Lax pair (\ref{l1})-(\ref{l2}) at $\Lambda = \Lambda_k $. Now using the notation $\Theta^{(k)} = \Theta \Lambda^{-k} $, $\Theta[0]=\Theta_1 $, $\psi[0]=\psi $ and for $N\geq 1 $ we write
\bea
\psi[N] &=& \psi[N-1]-\Theta^{(1)}[N-1] \left( \Theta[N-1] \right)^{-1} \psi^{(1)}[N-1], \nonumber \\
&=&\left\vert
\begin{array}{cccc}
  \Theta_1 & \cdots & \Theta_N & \psi \\
  \vdots & \ddots & \vdots & \vdots \\
  \Theta_1^{(N-1)} & \cdots & \Theta_N^{(N-1)} & \psi^{(N-1)} \\
  \Theta_1^{(N)} & \cdots & \Theta_N^{(N)} & \frame{\fbox{$ \psi^{(N)}$ }}
\end{array}
\right\vert .
\eea
Similarly the expression for $S[N] $ is
\bea
S[N]&=& S - \sum_{j=1}^{N-1} \Theta[j]^{(1)} \left( \Theta[j] \right)^{-1} , \nonumber \\
&=&S+\left\vert
\begin{array}{cccc}
  \Theta_1 & \cdots & \Theta_N & O \\
  \vdots & \ddots & \vdots & \vdots \\
  \Theta_1^{(N-2)} & \cdots & \Theta_N^{(N-2)} & O \\
  \Theta_1^{(N-1)} & \cdots & \Theta_N^{(N-1)} & I \\
  \Theta_1^{(N)} & \cdots & \Theta_N^{(N)} & \frame{\fbox{$ O $}}
\end{array}
\right\vert .
\eea
These results can be proved by induction and the proof is identical to the one for noncommutative KP equation (see e.g. \cite{gr3}-\cite{gilsonb} ).

\section{The $SU(2)$ system}

In order to get explicit solutions and compare our results with the already known soliton solutions of the coupled dispersionless integrable system, we proceed as follows. Let us make general remarks about the case when $\g=SU(n) $ and then focus on the specific case of $\g=SU(2) $. The Lie group $\g =SU(n)$ consists of unimodular anti-hermitian $n \times n $ matrices. Since the matrix fields $S$ and $G$ are valued in the Lie algebra ${\bf su}(n)$ of the Lie group $SU(n)$, therefore,
\bea
S^\dag &=& -S , \quad \quad \quad G^\dag \, =\, -G,\nonumber \\
{\rm Tr }S&=& 0, \quad \quad \quad \quad{\rm Tr }G\,=\, 0.
\eea
In order to have $S[1]$ and $G[1]$ to be valued in ${\bf su}(n)$, we need to have $M^\dag =-M $ and ${\rm Tr}M=0 $. For the particular solutions $\theta_i $ at $\lambda = \lambda_i $, let us compute
\bea
\del_x (\theta_i ^\dag \theta_j )&=& \bar{\lambda}_i \theta_i^\dag \del_x S^\dag \theta_j +\lambda_j \theta_i^\dag \del_x S \theta_j , \nonumber \\
\del_x (\theta_i ^\dag \theta_j )&=& \theta_i^\dag \left([S, G]^\dag +[S, G]\right)\theta_j+\bar{\lambda}_i ^{-1} \theta_i^\dag G^\dag \theta_j + \lambda_j^{-1}\theta_i^\dag G  \theta_j.
\eea
Since $S$ and $G$ are anti-hermitian, therefore, we get
\be
\del_x \left( \theta_i^\dag \theta_j \right)= \del_t \left( \theta_i^\dag \theta_j \right)=0,
\ee
when $\lambda_i \neq \lambda_j ( ${\rm i.e.}$ \bar{\lambda}_i = \lambda_j )$. Also from the definition of the the matrix $M$, we can calculate
\be
\theta_i^\dag \left(M^\dag + M \right) \theta_j = \left(\bar{\lambda}_i ^{-1} + {\lambda}_j ^{-1} \right) \theta_i^\dag \theta_j ,
\ee
which implies
\be
\theta_i^\dag \theta_j =0 ,\label{12}
\ee
when $\lambda_i \neq \lambda_j $. Given this, we conclude that the column vectors $\theta_i $ are all linearly independent and the condition (\ref{12}) holds at every point of space and time.

For the system with $\g = SU (2)$, the corresponding equation of motion consists of the hierarchy of localized induction equation of a thin vortex filament which is equivalent to the Heisenberg spin equation and the hierarchy of dispersionless equations which are equivalent to Pohlmeyer-Lund-Regge system and the sine-Gordon equation. Here we shall consider the coupled dispersionless integrable system which is equivalent to the sine-Gordon equation \cite{konno}. Let us introduce a vector ${\bf \phi}=(\phi^1 , \phi^2 , \phi^3 ) $ in such a way that the matrix field $S$ is defined by
\be
S= \ti \left(
       \begin{array}{cc}
         \phi^3 & \phi^1 - \ti \phi^2 \\
         \phi^1 + \ti \phi^2 & -\phi^3 \\
       \end{array}
     \right),
\ee
which is traceless and anti-hermitian. The matrices $U$ and $V$ are then given by
\bea
U&=& \ti\lambda \left(
              \begin{array}{cc}
                \del_x \phi^3 & \del_x\phi^1 - \ti \del_x\phi^2  \\
                \del_x\phi^1 + \ti \del_x \phi^2 & -\del_x\phi^3 \\
              \end{array}
            \right) ,\nonumber \\
V&=&\left(
    \begin{array}{cc}
      0 & \phi^1 - \ti \phi^2 \\
      -\phi^1 - \ti \phi^2 & 0 \\
    \end{array}
  \right)-\frac{\ti}{2\lambda} \left(
                               \begin{array}{cc}
                                 1 & 0 \\
                                 0 & -1 \\
                               \end{array}
                             \right).
\eea
By writing $\phi^1 =r ,\, \phi^2=0, \,  \phi^3=q $, we get the coupled dispersionless integrable system of ref. \cite{konno}
\bea
\del_x\del_t q + 2 \del_x r r&=&0 ,\nonumber \\
\del_x \del_t r -2 \del_xqr &=& 0 .\label{cd}
\eea
The soliton solutions to this system have been computed in \cite{chen} by using Backlund and Darboux transformations. In ref. \cite{chen}, the Darboux matrix $D(\lambda)$ has been obtained using the method described in the section 3 and is given by
\be
D(\lambda)= \left(
              \begin{array}{cc}
                \lambda^{-1}- \lambda_1^{-1} \cos \omega & -\lambda_1^{-1}\sin\omega \\
                -\lambda_1^{-1}\sin\omega & \lambda^{-1}+ \lambda_1^{-1} \cos \omega  \\
              \end{array}
            \right),
\ee
where $\tan \frac{\omega}{2} =\frac{\beta}{\alpha} $ and $(\alpha , \beta )^T $ is a particular column solution of the Lax pair at $\lambda=\lambda_1 $. The corresponding matrix $M$ is
\be
M=\lambda_1^{-1}\left(
    \begin{array}{cc}
      \cos \omega & \sin\omega \\
      \sin\omega & -\cos \omega \\
    \end{array}
  \right),
\ee
which is traceless and is anti-hermitian when $\bar{\lambda}_1=-{\lambda}_1 $. The one-fold Darboux transformation (\ref{dt1}) gives the Darboux transformation on the fields $q$ and $r$
\bea
q[1]&=& q+\ti \lambda_1^{-1}  \cos \omega ,\nonumber \\
r[1]&=& r +\ti \lambda_1^{-1} \sin\omega .
\eea
By assuming the seed solutions to be $\del_x q=1 $ and $r=0$, the one-soliton solutions turn out to be
\bea
\del_x q[1]&=& 1 +2 {\rm sech}^2 \left(2 \ti \lambda_1 x - \frac{\ti }{\lambda_1 } t \right), \nonumber \\
r[1]&=& \frac{\ti }{\lambda_1 }{\rm sech} \left(2 \ti \lambda_1 x - \frac{\ti }{\lambda_1 } t \right),
\eea
where $\del_x q[1] $ is called a soliton of dark-type and $r[1]$ is called a soliton of bright-type \cite{konno}.

To get simpler expressions of Darboux transformation on scalar solutions (eigenfunctions) of the Lax pair and the scalar fields $q$ and $r$, we make a gauge transformation on matrix fields $S$, $U$ and $V$. Let us introduce a matrix
\be
\Omega=\frac{1}{\sqrt{2}}\left(
     \begin{array}{cc}
       1 & 1 \\
       -\ti & \ti \\
     \end{array}
   \right),
\ee
so that the gauge equivalent matrix field $\widetilde{S}$ is given by
\be
\widetilde{S}=\Omega^{-1} S \Omega =\ti\left(
                            \begin{array}{cc}
                              0 & q+\ti r \\
                              q-\ti r & 0 \\
                            \end{array}
                          \right).
\ee
The Lax pair is then expressed as
\bea
\del_x \left(
         \begin{array}{c}
           X \\
           Y \\
         \end{array}
       \right)
&=& \ti \lambda \left(
                \begin{array}{cc}
                  0 & \del_x (q+\ti r) \\
                  \del_x (q-\ti r) & 0 \\
                \end{array}
              \right)
\left(
  \begin{array}{c}
    X \\
    Y \\
  \end{array}
\right), \nonumber \\
\del_t \left(
         \begin{array}{c}
           X \\
           Y \\
         \end{array}
       \right)
&=&  \left(
                \begin{array}{cc}
                  -\ti r & -\frac{\ti}{2\lambda} \\
                  -\frac{\ti}{2\lambda} & \ti r \\
                \end{array}
              \right)
\left(
  \begin{array}{c}
    X \\
    Y \\
  \end{array}
\right). \label{lax2}
\eea
Let $X_1 $ and $Y_1 $ be the particular scalar solutions of the Lax pair (\ref{lax2}) at $\lambda=\lambda_1 $ then the one-fold Darboux matrix for the Lax pair (\ref{lax2}) is given by
\be
\widetilde{D}(\lambda )=\left(
                          \begin{array}{cc}
                            \lambda^{-1} & - \lambda_1^{-1} \frac{X_1}{Y_1} \\
                            -\lambda_1^{-1} \frac{Y_1}{X_1} & \lambda^{-1} \\
                          \end{array}
                        \right).
\ee
In other words the Lax pair (\ref{lax2}) is covariant under the following Darboux transformations
\bea
X[1]&=& X^{(1)}- \frac{X_1^{(1)}}{ Y_1} Y \,=\, \frac{\Delta_1 (X_1 , Y_1 , X, Y )[2]}{\Delta_2 (X_1 , Y_1)[1] }, \nonumber \\
Y[1]&=& Y^{(1)}- \frac{Y_1^{(1)}}{X_1} X^{(1)} \,=\, \frac{\Delta_2 (X_1 , Y_1 , X, Y )[2]}{\Delta_1 (X_1 , Y_1)[1] }, \nonumber \\
q[1]&=& q - \frac{\ti}{2} \left( \frac{X_1^{(1)}}{Y_1} + \frac{Y_1^{(1)}}{X_1} \right)\,=\,q + \del_t \log \left(\Delta_1 (X_1 , Y_1)[1] \Delta_2 (X_1 , Y_1)[1]\right), \nonumber \\
r[1]&=&r - \frac{1}{2} \left( \frac{X_1^{(1)}}{Y_1}
-\frac{Y_1^{(1)}}{X_1} \right) \,=\, -r + \ti \del_t \log
\left(\frac{\Delta_1 (X_1 , Y_1)[1]}{\Delta_2 (X_1 , Y_1)[1]}
\right), \eea where we have used the notation
$X^{(1)}=\lambda^{-1} X$ and the determinants $\Delta_1 $ and
$\Delta_2 $ are defined by \bea \Delta_1 (X_1 , Y_1 , X, Y
)[2]&=&\left \vert\begin{array}{cc}
                      Y_1 & Y \\
                      X^{(1)}_1 & X^{(1)}
                    \end{array}\right \vert , \quad \Delta_2 (X_1 , Y_1 , X, Y )[2]\,=\, \left \vert\begin{array}{cc}
                      X_1 &  X\\
                      Y_1^{(1)} & Y^{(1)}
                    \end{array}\right \vert ,\nonumber \\
\Delta_1 (X_1 , Y_1)[1]&=& X_1 , \hspace{1in} \quad \quad \quad   \Delta_2 (X_1 , Y_1)[1]\,=\, Y_1 .\nonumber
\eea
Similarly the two-fold Darboux transformations on scalar solutions $X$, $Y$ of the lax pair (\ref{lax2}) and  $q$, $r$ of the coupled dispersionless integrable system (\ref{cd}) are given by
\bea
X[2]&=&X^{(1)}[1]- \frac{X_1^{(1)}[1]}{ Y_1[1]} Y[1] \,=\, \frac{\Delta_1 (X_k , Y_k , X, Y )[3]}{\Delta_2 (X_k , Y_k)[2] }, \nonumber \\
Y[2]&=& Y^{(1)}- \frac{Y_1^{(1)}}{X_1} X^{(1)} \,=\, \frac{\Delta_2 (X_1 , Y_1 , X, Y )[3]}{\Delta_1 (X_1 , Y_1)[2] }, \nonumber \\
q[2]&=& q [1] - \frac{\ti}{2} \left( \frac{X_1^{(1)}[1]}{Y_1[1]} + \frac{Y_1^{(1)}[1]}{X_1[1]} \right)\,=\,q + \del_t \log \left(\Delta_1 (X_k , Y_k)[2] \Delta_2 (X_k , Y_k)[2]\right), \nonumber \\
r[2]&=&r[1] - \frac{1}{2} \left( \frac{X_1^{(1)}[1]}{Y_1[1]}
-\frac{Y_1^{(1)}[1]}{X_1[1]} \right) \,=\, r + \ti \del_t \log
\left(\frac{\Delta_1 (X_k , Y_k)[2]}{\Delta_2 (X_k , Y_k)[2]}
\right), \eea where for each $k=1,2 $ we have assumed $X_k$ ,
$Y_k$ to be particular solutions of the Lax pair (\ref{lax2}) at
$\lambda =\lambda_k $ and the determinants $\Delta_1 $ and
$\Delta_2 $ are defined by \bea \Delta_1 (X_k , Y_k , X, Y
)[3]&=&\left \vert \begin{array}{ccc}
                                                X_1 & X_2 & X \\
                                                Y_1^{(1)} & Y_2^{(1)} & Y^{(1)} \\
                                                X_1^{(2)} & X_2^{(2)} & X^{(2)}
                                              \end{array} \right \vert ,\nonumber \\
\Delta_2 (X_k , Y_k , X, Y )[3]&=&\left \vert \begin{array}{ccc}
                                                Y_1 & Y_2 & Y \\
                                                X_1^{(1)} & X_2^{(1)} & X^{(1)} \\
                                                Y_1^{(2)} & Y_2^{(2)} & Y^{(2)}
                                              \end{array} \right \vert , \nonumber \\
\Delta_1 (X_k , Y_k)[2]&=&\left \vert\begin{array}{cc}
                      Y_1 & Y_2 \\
                      X^{(1)}_1 & X^{(1)}_2
                    \end{array}\right \vert , \nonumber \\
\Delta_2 (X_k , Y_k)[2]&=&\left \vert\begin{array}{cc}
                      X_1 & X_2 \\
                      Y^{(1)}_1 & Y^{(1)}_2
                    \end{array}\right \vert .
\eea
We can iterate the Darboux transformation $N$ times to arrive at the $N$-soliton solutions to the coupled dispersionless integrable system (\ref{cd}) expressed in closed form as ratios of determinants. For each $k=1,\cdots, N $, let $X_k $ and $Y_k $ be $N$ particular scalar solutions of the Lax pair at $\lambda=\lambda_k$ respectively then by using the notation $X_j^{(k)}=\lambda_j^{-k}X_j $ and $X^{(k)}=\lambda^{-k}X $, we get the following result
\bea
X[N]&=&\frac{\Delta_1 (X_k , Y_k , X,Y)[N+1]}{\Delta_2 (X_k , Y_k)[N]}, \nonumber \\
Y[N]&=&\frac{\Delta_2 (X_k , Y_k , X,Y)[N+1]}{\Delta_1 (X_k , Y_k)[N]},
\eea
where the determinants are defined by the following expressions. For $N$ odd the determinants $\Delta_1 $ and $\Delta_2 $ are given by
\bea
\Delta_1 (X_k , Y_k , X,Y)[N+1]&=& \left \vert \begin{array}{cccc}
                            Y_1 & \cdots & Y_N & Y \\
                            \vdots & \ddots & \vdots & \vdots \\
                            Y^{(N-1)}_1 & \cdots & Y^{(N-1)}_N & Y^{(N-1)} \\
                            X^{(N)}_1 & \cdots & X_N^{(N)} & X^{(N)}
                          \end{array} \right \vert , \nonumber \\
\Delta_2 (X_k , Y_k , X,Y)[N+1]&=& \left \vert \begin{array}{cccc}
                            X_1 & \cdots & X_N & X \\
                            \vdots & \ddots & \vdots & \vdots \\
                            X^{(N-1)}_1 & \cdots & X^{(N-1)}_N & X^{(N-1)} \\
                            Y^{(N)}_1 & \cdots & Y_N^{(N)} & Y^{(N)}
                          \end{array} \right \vert .
\eea
For $N$ even we have the following determinants
\bea
\Delta_1 (X_k , Y_k , X,Y)[N+1]&=& \left \vert \begin{array}{cccc}
                            X_1 & \cdots & X_N & X \\
                            \vdots & \ddots & \vdots & \vdots \\
                            Y^{(N-1)}_1 & \cdots & Y^{(N-1)}_N & Y^{(N-1)} \\
                            X^{(N)}_1 & \cdots & X_N^{(N)} & X^{(N)}
                          \end{array} \right \vert , \nonumber \\
\Delta_2 (X_k , Y_k , X,Y)[N+1]&=& \left \vert \begin{array}{cccc}
                            Y_1 & \cdots & Y_N & Y \\
                            \vdots & \ddots & \vdots & \vdots \\
                            X^{(N-1)}_1 & \cdots & X^{(N-1)}_N & X^{(N-1)} \\
                            Y^{(N)}_1 & \cdots & Y_N^{(N)} & Y^{(N)}
                          \end{array} \right \vert .
\eea
Similarly after performing Darboux transformations $N$ times, we get the following expressions of the $N$-soliton solutions to the couple dispersionless integrable system (\ref{cd})
\bea
q[N]&=& q+\del_t \log \left( \Delta_1 (X_k , Y_k )[N]  \Delta_2 (X_k , Y_k )[N] \right) ,\nonumber \\
r[N]&=& (-1)^N r+ \ti \del_t \log \left( \frac{\Delta_1 (X_k , Y_k )[N]}{\Delta_2 (X_k , Y_k )[N]}  \right).
\eea

Since the system (\ref{cd}) is equivalent to the sine-Gordon system, therefore, the solutions to both the system are related with each other. The natural boundary conditions for the system (\ref{cd}) are
\be
\del_x q \rightarrow 1\,\,({\rm const. }), \quad \quad r\rightarrow 0, \quad \quad |x|\rightarrow\infty . \nonumber
\ee
With these boundary conditions, the solution $\varphi $ to the sine-Gordon equation $\del_x \del_t \varphi = 2\sin \varphi $ and the solutions $q$ and $r$ to the system (\ref{cd}) are related by
\be
\del_x q =\cos \varphi , \quad \quad \quad r=\pm \frac{1}{2} \del_t \varphi . \nonumber
\ee
The $N$-fold Darboux transformation on $\varphi $ is given by
\bea
\varphi[N]&=& \varphi +2 \ti \log \left( \frac{\Delta_1 (X_k , Y_k )[N]}{\Delta_2 (X_k , Y_k )[N]} \right), \nonumber \\
\exp \left(- \ti \varphi [N] \right)&=& \exp \left(- \ti \varphi \right) \left( \frac{\Delta_1 (X_k , Y_k )[N]}{\Delta_2 (X_k , Y_k )[N]} \right)^2 ,
\eea
which are the well-known expressions of $N$-soliton solutions to the sine-Gordon equation \cite{matveev}.
\section{Concluding remarks}

In this paper, we have studied the Darboux transformation on matrix solutions to the generalized coupled dispersionless integrable system based on some non-abelian Lie group and expressed the matrix solutions in terms of quasideterminants. For the particular case of the coupled dispersionless system which is equivalent to the sine-Gordon equation, we have defined a Darboux transformation on scalar solutions of the linear problem and expressed them as ratios of determinants as is the case with the sine-Gordon equation. There are a number of directions where the present work could be extended and the immediate one is to study the effect of Moyal noncommutativity of space coordinates on the soliton solutions of the system. The other interesting direction to pusue is to study the supersymmetric generalization of the generalized coupled dispersionless integrable system. We shall address these and related problems in some later work.

{\large {\bf {Acknowledgements}}}

I thank Jonathan Nimmo for helpful discussions and the Department of Mathematics,
University of Glasgow, for warm hospitality and provision of facilities while this
work was done. I gratefully acknowledge financial support of the Higher Education
Commission of Pakistan.



\bigskip

\end{document}